\documentclass[aps,prl,floatfix,twocolumn,footinbib,amsmath,amssymb]{revtex4}
\usepackage{amsmath,amsthm,amssymb,color,psfrag}
\usepackage{url}
\usepackage{latexsym}
\usepackage{graphicx}

\definecolor{darkred}{rgb}{0.6,0.0,0.0}
\definecolor{darkblue}{rgb}{0.0,0.0,0.5}
\definecolor{darkgreen}{rgb}{0.0,0.5,0.0}
\definecolor{brown}{rgb}{0.0,0.0,0.0}
\newcommand{\red}{\color{darkred}}
\newcommand{\blue}{\color{darkblue}}

\begin{document}

\title{Quark and Gluon Tagging at the LHC}
\author{Jason Gallicchio}
\author{Matthew D. Schwartz}
\affiliation{Department of Physics
Harvard University
Cambridge, MA 02138, U.S.A}

\begin{abstract}
Being able to distinguish light-quark jets from gluon jets on an
event-by-event basis could significantly enhance the reach for many new physics searches
at the Large Hadron Collider.
Through an exhaustive search of existing
and novel jet substructure observables, we find that
a multivariate approach can
filter out over 95\% of the gluon jets while
keeping more than half of the light-quark jets. Moreover, a
combination of two simple variables, the charge track multiplicity and
the $p_T$-weighted linear radial moment (girth), can achieve similar results.
While this pair appears very promising, our study is only Monte
Carlo based, and other discriminants may work better with real
data in a realistic experimental environment. To that end, we
explore many other observables constructed using different jet
sizes and parameters, and highlight those that deserve further
theoretical and experimental scrutiny. Additional information,
including distributions of around 10,000 variables, can be
found on the website \url{http://jets.physics.harvard.edu/qvg}.
\end{abstract}
\maketitle

The Large Hadron Collider at CERN produces billions of
jets a second. Understanding these jets may be the
key to unraveling physics beyond the standard model. The
jets at the LHC can be coarsely partitioned into quark
or gluon jets. Almost always, the jets we would like most
to study are quark jets and so removing the gluon
jet background, if possible, could greatly enhance the
search reach for many new physics scenarios.
In this letter, we show that through a comprehensive
study of light quark (uds) vs gluon discriminants, many of which take
advantage of the improved resolution of the LHC detectors,
gluon-tagging with good efficiency is achievable on
an event-by-event basis.

There are many situations in
which quark/gluon tagging would be useful.
For example, the jets produced in supersymmetric decay
chains
are usually entirely quark jets, while their backgrounds are mostly
gluon jets.
Quark tagging would be especially
helpful in cases like R-parity violating SUSY where there are
no additional handles like leptons, photons or missing energy.
Even when there are leptonic decay modes, quark tagging
could help measure the hadronic branching ratio of new physics,
which will be needed to verify the model.
Interesting standard
model physics also tends to be quark-heavy. For example,
vector-boson fusion requires
the tagging of two forward jets which are always quarks,
while background jets in this forward region tend to be dominated
by gluons. Alternatively, there are some scenarios where the new physics
is in gluon jets (e.g. coloron models~\cite{Kilic:2008ub}, which produce 4
or more gluon jets), for which gluon tagging would also help.

Quark and gluon jets are an extremely useful abstraction, discussed in hundreds of papers and many experimental
studies, despite their not having a precise theoretical or experimental definition~\cite{Banfi:2006hf}.
Any flavor tagging is only meaningful to the extent that there is a
correspondence between hard partons and jets, which is the standard starting point
for almost every collider search involving hadronic final states. The
correspondence  is affected by things like the jet algorithm,
the event's topology, and the distance between energetic deposits.
In this paper, quark or gluon jets refer to the parton which is produced
in the hard process at leading order in perturbation theory and initiates the parton shower.
At this level there is no ambiguity in what is meant by the jet flavor since there
is no interference between different final states.
In fact, the flavor is well-defined to to all orders in perturbation theory up to the same power corrections
that affect any collinear and IR safe jet algorithm's parton correspondence~\cite{Gallicchio:2011xc}.
These power corrections involve the jet size $R$ (equivalently
the jet's mass-to-energy ratio $m/E$) and $\Lambda_{\mathrm{QCD}}/E$.

In this paper, we study pure quark or gluon samples and
only consider observable properties of these jets. The expectation is that properties of jets
coming from, say, SUSY decays will be more similar to the properties of the quark jets in our simulation,
despite the fact that jet properties are not expected to be completely universal --  jet
substructure is affected by things like adjacent jets and differing color-connections.
The result of this paper is a rough ranking of jet observables that can be used to distinguish
light flavor and help find new physics signals.   Once promising observables are found and
their distributions measured, it might be possible to construct an even more powerful tagger.

Gluon tagging at the LHC is both more useful and achievable
than at the Tevatron. The LHC's proton-proton initial state,
higher energy, and higher luminosity increase the number of
jets produced and make gluon jet backgrounds more common. CMS's
particle flow \cite{particleflow} and {\sc Atlas}'s
individually calibrated TopoClusters \cite{Lampl:2008zz}, along
with each detector's improved calorimeter resolution, allow
unprecedented measurement of the energy and track distribution
within jets.
As we will see, the better the resolution on the jet constituents,
the better the tagger will be.
We also find that
the higher $p_T$ jets of the LHC are more reliably tagged than
lower $p_T$ jets of previous colliders, as long as the tracking
remains reliable.

Much is known experimentally about quark and gluon jets (see \cite{Dissertori:2003} for a summary).
One important LEP result was that $b$-jets were more similar to
gluon jets than to light-quark jets
\cite{Buskulic:1995sw,Biebel:1996mc}: due to the longer decay
chain of $B$-hadrons, the number of particles and
angular spread is larger for a $b$-jet than a light-quark jet.
The similarity of $b$-jets to gluon jets should be lessened in
the LHC's higher $p_T$ jets because the QCD shower produces
more particles, whereas the particle multiplicity is relatively
fixed in the $B$-hadron decay. There are already
sophisticated and very detector-specific methods for $b$-tagging.
Current $b$-taggers rely mostly on impact parameters or a
secondary vertex, so they are independent of the observables we
consider. Therefore, we restrict our study to discriminating light quarks (uds)
from gluons.

The accumulated knowledge from decades of experiments
and perturbative QCD calculations have been incorporated into
Monte Carlo event generators, in particular Pythia~\cite{Sjostrand:2007gs}
and Herwig~\cite{Bahr:2008pv}.
These programs also include sophisticated hadronization
and underlying event models which have also been tuned to data.
Small differences still exist between these tools (and between
the tools and data), but they provide an excellent starting point
to characterize which observables might be useful in gluon-tagging.
The approach to gluon-tagging discussed here is to find observables
which appear promising and then can be measured and calibrated
on samples of mixed or pure quark or gluon jets at the LHC~\cite{Gallicchio:2011xc}.

\begin{figure}[t]
   \psfrag{shape_integ_pT_R_jet0_ak07_particles_bin_20_of_38_0100}{$\Psi(r=0.1)$ for 200\,GeV Jets}
   \psfrag{Q 200 GeV}{\blue{Quark}}
   \psfrag{G 200 GeV}{\red{Gluon}}
   \includegraphics[width=0.4\textwidth]{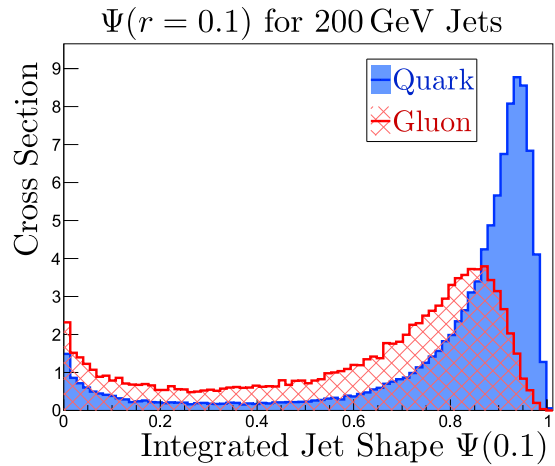}
\caption{
Data on the integrated jet shape $\Psi(r)$ is usually published only when averaged over all events.
Here we show the distribution of $\Psi(0.1)$, for quarks (blue, solid) and gluons (red, hollow).
The event-by-event distributions of $\Psi(r)$ and other
observables are much more important for gluon tagging than average values.
} \label{fig:jetshape}
\end{figure}

To understand the structure of a jet, it is important to distinguish
observables which average over all events from observables
which are useful on an event-by-event basis.
One example of an averaged observable is the classic {\it integrated jet
shape}, $\Psi(r)$, which has already been measured at the LHC~\cite{Aad:2011kq}.
This jet shape is defined as the
fraction of a jet's $p_T$ within a cone of radius $r$.
Traditionally, jet shapes are presented as an average
over all jets in a particular $p_T$ or $\eta$ range. For any
$r$, the averaged jet shape becomes a single number,
which is generally larger for quarks than for gluons
because a greater fraction of a typical quark jet's $p_T$ is
at the center of the jet.
On traditional jet shape plots, error bars for each $r$ are proportional to the standard deviation of
the underlying distribution, but that distribution is \emph{not} a narrow Gaussian
around the average. For example, the event-by-event
distributions for $\Psi(r=0.1)$ are shown in
Figure~\ref{fig:jetshape} for quarks and gluons.
Jet shapes averaged over these distributions throw out useful information about the location
and $p_T$'s of particles within the jet, along with their
correlations.
For event-by-event discrimination, it is crucial to have
distributions, whereas most public data only describes averages.
In this study we consider $\Psi(r)$ and
many other variables to see which are best suited to quark/gluon tagging.

To generate samples of quark and gluon jets we considered samples
of dijet events and $\gamma$+jet events. These
were generated with {\sc madgraph
v4.4.26}~\cite{Alwall:2007st} and showered through both {\sc
pythia v8.140}~\cite{Sjostrand:2007gs} and {\sc herwig++
v2.4.2}~\cite{Bahr:2008pv} with the default tunes. Jets,
reconstructed using {\sc fastjet v2.4.2}~\cite{Cacciari:2005hq},
were required to have $|\eta|<1$.

We needed to isolate samples of quark and gluon jets with the
similar jet $p_T$'s. Unfortunately, we cannot get similar
jet $p_T$'s by having similar $p_T$'s at the hard parton level,
since the showering changes the $p_T$ significantly, and differently
for quarks and gluons. This is an unphysical difference, since the parton
$p_T$ is set artificially, and we have to avoid our tagger picking up on it.
The solution
we chose was to generate and shower a wide spectrum of dijet and
$\gamma$+jet events, and require the resulting Anti-$k_T$ R=0.5
jets to lie within 10\% of the central value for each of six
$p_T$ windows, centered around 50, 100, 200, 400, 800, and
1600\,GeV. (The underlying hard partons spanned a range from
half to twice the central value.) The $p_T$ spectrum within
each window matches the falling spectrum of the underlying
dijet or $\gamma$+jet samples, which are nearly
identical for quarks and gluons in narrow windows chosen. When
the entire event is reclustered with a different jet size, as
was done when examining how the observables change with $R$,
the resulting jet $p_T$ no longer necessarily lies within the
narrow $\pm$10\% window.  In fact, how the jet $p_T$ changes
with $R$ forms a quark/gluon discriminant similar to integrated jet shape.

With each sample of similar-$p_T$ jets, there are two main
types of observables useful in separating quarks from gluons:
{\it discrete} ones, which try to distinguish individual
particles/tracks/subjets, and {\it continuous} ones that can treat the energy or
$p_T$ within the jet as a smooth function of $(\delta \eta,
\delta \phi)$ away from the jet axis in order to form
combinations like geometric moments.

The discrete category includes the number of distinguishable tracks, small
subjets, or reconstructed particles. Functions of this information,
such as the average or the spread (standard deviation) of their $p_T$'s were considered.
We find that this class of
observables provide the best discrimination at high quark
efficiency (mild cuts) and high jet $p_T$.


The average multiplicity
of \emph{any} type of particle, along with its variance, are
sensitive to the QCD charges of the underlying gluon ($C_A=3$)
 or quark ($C_F=4/3$). To leading order,
\begin{equation}
    \frac{\langle N_g \rangle}{\langle N_q \rangle} = \frac{C_A}{C_F}
    \qquad
    \textrm{and}
    \qquad
    \frac{\sigma_g^2}{\sigma_q^2} = \frac{C_A}{C_F}.
\end{equation}
The OPAL collaboration, among others, studied the charged
particle multiplicity in light-quark and gluon jets of energy
around 40\,GeV to 45\,GeV \cite{Ackerstaff:1997xg} and found
distributions that agree well with the Monte Carlo event
generators and with analytic predictions.

\begin{figure*}
\psfrag{Signal}{\qquad \qquad \qquad \blue{Quark}}
\psfrag{components_jet0_ak05_charged_count}{\small{\!\!\!\!\!\!\!\!\!\!Charged Track Multiplicity}}
\psfrag{components_jet0_ak07_charged_count}{\small{\!\!\!\!\!\!\!\!\!\!Charged Track Multiplicity}}
\psfrag{shape_optimal_pT_R_jet0_ak04_particles}{\small{Optimal Shape}}
\psfrag{radial_moment_ak05_1_Pt_4v_y}{\small{\!\!\!\!\!\!\!\!\!\!\!Linear Radial Moment}}
\psfrag{radial_moment_ak04_1_Pt_4v_y}{\small{\!\!\!\!\!\!\!\!\!\!\!Linear Radial Moment}}
\includegraphics[width=0.3\textwidth]{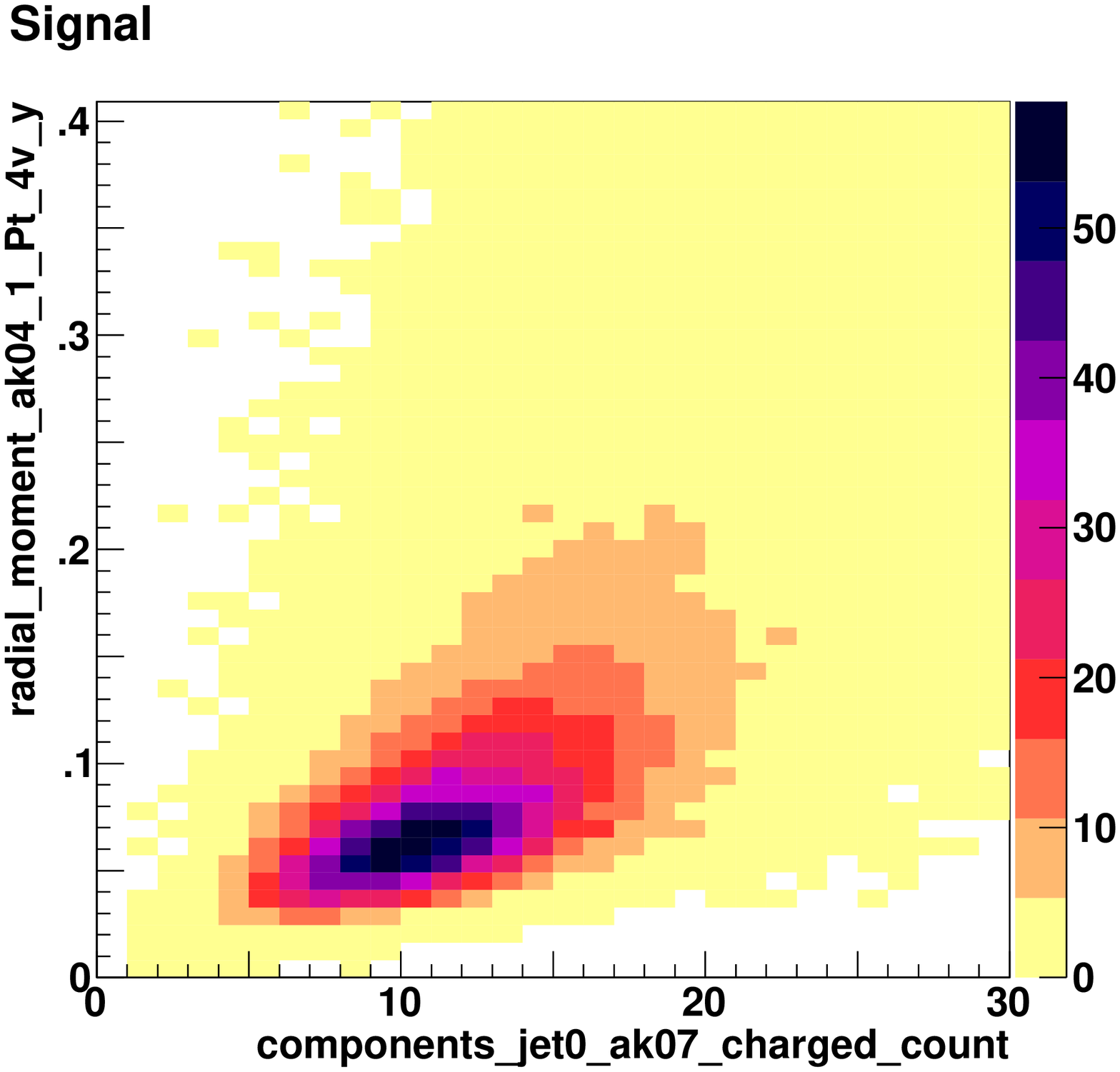}
\psfrag{Background}{\qquad \qquad \qquad \red{Gluon}}
\includegraphics[width=0.3\textwidth]{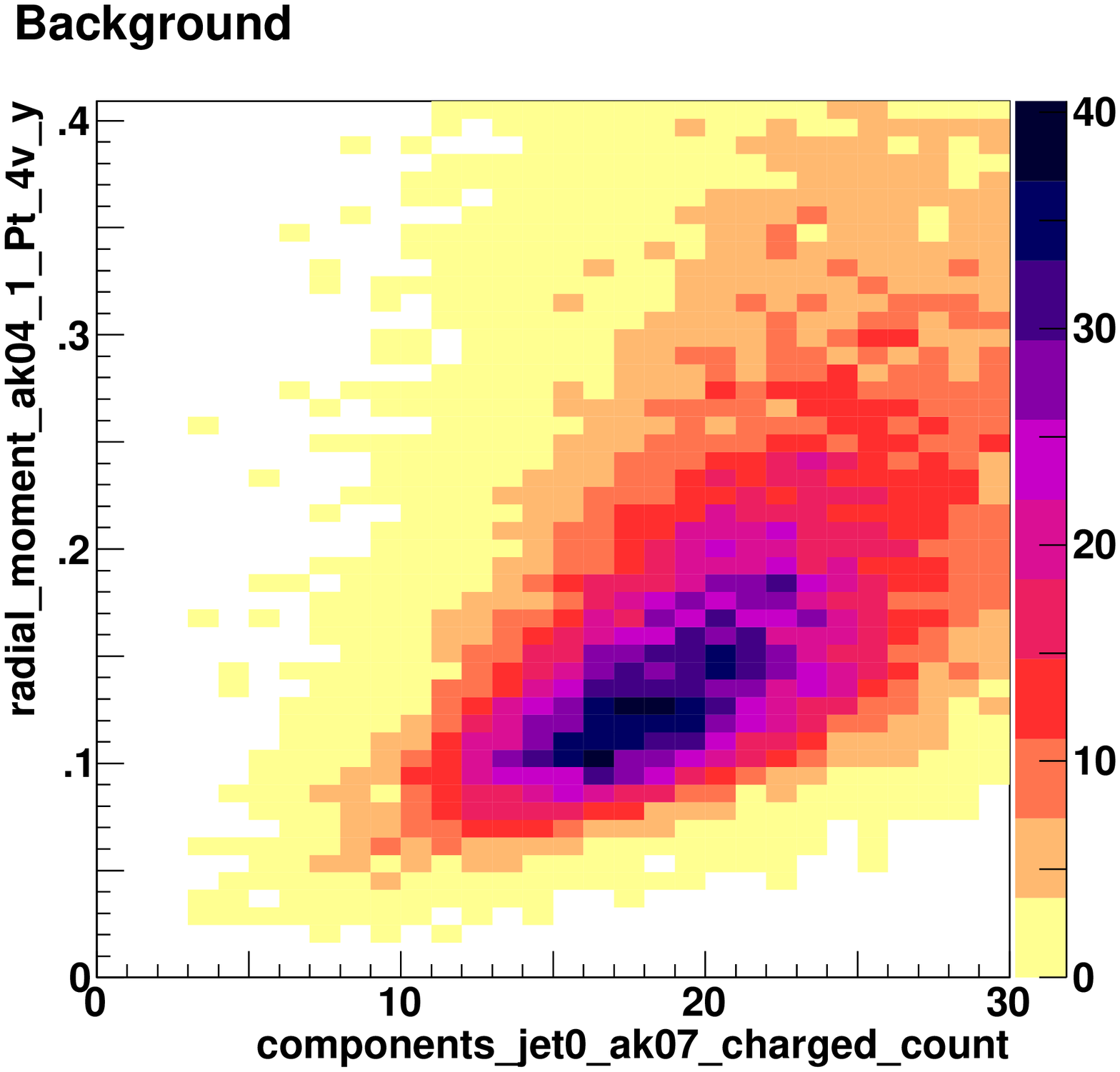}
\psfrag{Likelihood}{\qquad \qquad Likelihood: $q/(q+g)$}
\psfrag{components_jet0_ak05_charged_count_shape_optimal_pT_R_jet0_ak04_particles}{\qquad \scriptsize{Log(Q/G)}}  
\includegraphics[width=0.3\textwidth]{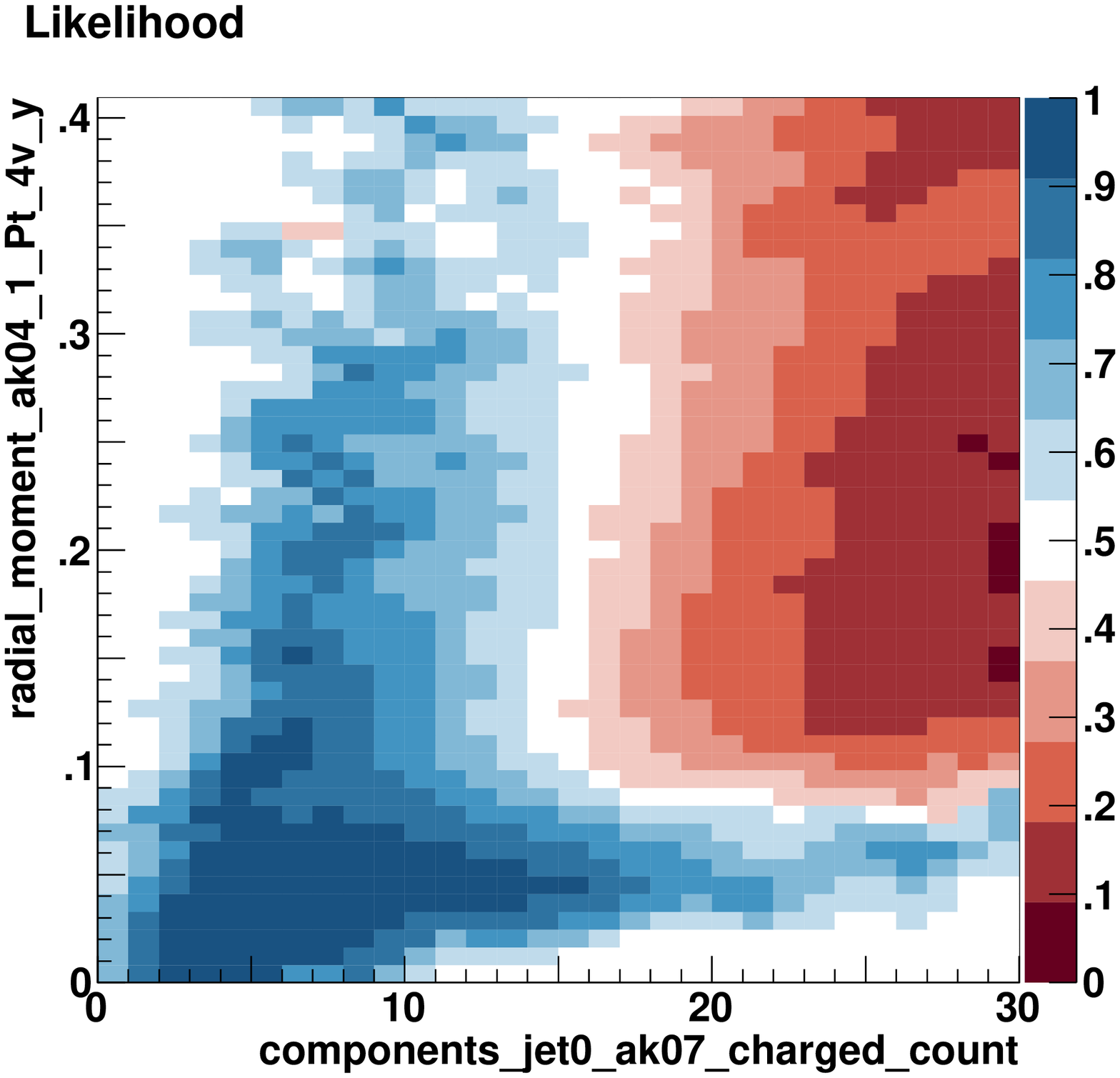}
\caption{ 2D Histograms of the two best observables, along with
the likelihood formed by combining them bin-by-bin. }
\label{fig:histo2d}
\end{figure*}

We find that the strongest discrete observable is the number of charged
particles within the jet, where charged particles were required to have
$p_T>500$\,MeV. Lower cuttoffs actually lead to better discrimination power,
so how well the LHC detectors will be able to resolve
particle $p_T$ will have important consequences for gluon tagging.

Another discrete observable is the subjet multiplicity, which was also
studied at LEP \cite{Barate:1998cp,Abreu:1998ve}. Extremely
small subjets approach the limit of particles and are sensitive to
hadronization, but larger subjets probe the better modeled,
perturbative physics \emph{and} give the largest ratio between
quark and gluon subjet multiplicities. For the higher-energy
jets of the LHC, the optimal jet size is far smaller than the
calorimeter resolution.  We found that counting
R$_\mathrm{sub}$=0.1 anti-$k_T$ jets was more powerful than
other subjet algorithms and larger sizes, but not as powerful
as counting the number of charged tracks. Small subjet
multiplicity can serve as a reasonable substitute if charged
track multiplicity proves less reliable in some circumstances
(perhaps at very high $\eta$).  Counting all hadrons works even better
than charged tracks.

Other observables in the discrete category that show
reasonable discrimination power include the average distance to jet
axis $\langle r \rangle$, the $p_T$ fraction of the
N$^\mathrm{th}$ hardest track or subjet, and the subjet
splitting scale (when the jet is reclustered with the $k_T$
algorithm).  Finally, there are observables that take advantage
of the electrical charge that quarks carry. Since the hardest
hadrons produced at the end of the shower have charges
correlated with the initiating quark, adding up the charges of
all tracks weighted by their $p_T$ gives some small
discrimination.


The second, more continuous, category of observables includes
jet mass, jet broadening \cite{Catani:1992jc}, and the family
of radial moments like girth~\cite{Gallicchio:2010dq},
angularities~\cite{Berger:2003iw}, and the optimal
moment which are described below. These tend to perform better
at lower jet $p_T$ and for achieving high quark purity through
harsh cuts. Other observables that try to capture the 2D shape
or color connections of the jet, like
pull~\cite{Gallicchio:2010sw}, eccentricity, or planar
flow~\cite{Almeida:2008yp}, are less powerful in this
application.

We find the best single observable in the continuous category
is the linear radial moment --
a measure of the `width' or `girth' of the jet -- constructed by adding up
the $p_T$ deposits within the jet, weighted by distance from
jet axis. It is defined as
\begin{equation}
    \textrm{Linear Radial Moment (Girth):}\quad  g =  \sum_{i \in \mathrm{jet}} \frac{ p_T^i }{ p_T^{jet} } \, |r_i|
\end{equation}
where $r_i = \sqrt{ \Delta y_i^2 + \Delta \phi_i^2 }$ and where
the true boost-invariant rapidity $y$ should be used for the
(massive) jet axis instead of the geometric pseudorapidity
$\eta$. Under the assumption of central jets with massless
constituents at small angles, this linear moment is identical
to jet broadening, defined as the sum of momenta transverse to
the jet axis normalized by the sum of momenta. While jet
broadening is natural at an $e^+e^-$ collider, the linear
radial moment is more natural and works a bit better at the
LHC. Other geometric moments involving different powers of $r$
were not as powerful at discriminating quarks from gluons,
including the jet mass, which is equal to the $r^2$ geometric
moment in the same limit.

By weighting the $p_T$ by other functions of $r$, whole
families of radial-kernel observables can be constructed:
\begin{equation}
    \textrm{Kernel Moment}:\qquad  K =  \sum_{i \in \mathrm{jet}} \frac{ p_T^i }{ p_T^{jet} } \, K(r_i)
\end{equation}
Angularities~\cite{Berger:2003iw,Almeida:2008yp} are one such
example, where the $p_T$ and $r$ are usually replaced by energy
and angle. Angularities are often normalized by the jet mass
rather than the jet $p_T$, and we considered both
normalizations. Both angularities and kernels which are powers
of $r$ suffer from sensitivity to the edge of the jet where
their kernels are greatest. This becomes problematic in crowded
environments with adjacent jets.

Rather than try to guess a useful kernel, we attempted to optimize
its shape numerically.
By parameterizing the kernel as a spline with 5 to 10 points, a
genetic algorithm was used to maximize gluon rejection for
several different quark efficiencies. In all cases, the optimal
kernels rose linearly from the axis of the jet out to $r \sim
0.3$, then turned over and decreased smoothly to zero at the
edge of the jet, but the gluon rejection in call cases was
rather insensitive to the region away from the center.
These optimal kernels performed slightly better than the linear
radial moments, but not enough to justify additional focus here.


By looking at combinations of observables, additional
quark/gluon discrimination is achieved. The 2D histograms for
the best discrete and continuous observables, charged particle
count and the linear radial moment, are shown in
Figure~\ref{fig:histo2d}. While the two observables are
correlated, it is still helpful to use both. In the third panel
of this figure, we show the 2D bin-by-bin likelihood
distribution. Given these variables, the discriminant that
achieves optimal gluon rejection for a fixed quark efficiency
is a simple cut on the appropriate likelihood contour. Cutting
out the top-right corner, for example, eliminates the most
egregiously gluey jets. In practice, this can be pre-computed
or measured in each jet $p_T$ window. As part of jet energy
scale calibrations, {\sc Atlas} \cite{Atlas:QvG} has measured
these two variables in dijet, $\gamma$-jet, and multijet
samples and used them individually to determine the flavor
composition to 10\% precision.

%
%
The same method can be applied for more than 2 observables, but
then the exact likelihood becomes impossible to map efficiently
with limited training samples.
A multivariate technique like Boosted Decision Trees can be
employed to approximate this multidimensional likelihood
distribution, as explained in~\cite{Gallicchio:2010dq}.

\begin{figure}[t]
      \psfrag{components_jet0_ak07_charged_count}{\scriptsize{charged mult R=0.5}}
      \psfrag{components_jet0_ak05_sub_ak010_count}{\scriptsize{subjet mult R$_\mathrm{sub}$=0.1}}
      \psfrag{components_jet0_ak07_sub_ak010_count}{\scriptsize{subjet mult R$_\mathrm{sub}$\!=0.1}}
      \psfrag{ak03_MDPt}{\scriptsize{mass/Pt R=0.3}}
      \psfrag{radial_moment_ak05_2cos_Pt_4v_y}{\scriptsize{cos moment R=0.5}}
      \psfrag{angularity_ak05_p100_Pt_4v_y_Rmax_overPt}{\scriptsize{ang a=1 R=0.5}}
      \psfrag{radial_moment_ak05_1_Pt_4v_y}{\scriptsize{girth R=0.5}}
      \psfrag{subjet_1stPt_over_jetPt_jet0_ak05_sub_ak010}{\scriptsize{1st subjet $p_T$}}
      \psfrag{decluster_kT_min_jet0_ak05_kt010}{\scriptsize{decluster $k_T$}}
      \psfrag{shape_optimal_pT_R_jet0_ak10_particles}{\scriptsize{optimal R=1.0}}
      \psfrag{pull_Eta_ak03_0_E_4v_y}{\scriptsize{pull $\eta$ R=0.3}}
      \psfrag{pull_R_ak03_2_Pt_4v_y}{\scriptsize{$|$pull$|$ R=0.3}}
      \psfrag{planar_flow_covariance_jet0_ak03_sub_ak010_kTE}{\scriptsize{planar flow R=0.3}}
      \psfrag{components_jet0_ak07_charged_count, radial_moment_ak05_1_Pt_4v_y}{\scriptsize{best pair}}
      \psfrag{group of length 5}{\scriptsize{group of 5}}
      \psfrag{combo005_ak02MDPt_componentsjet0ak04particlesbroadening_componentsjet0ak04subak010av_641d1850de95b29a}{\scriptsize{group of 5}}
      \psfrag{components_jet0_ak07_charged_count, radial_moment_ak04_1_Pt_4v_y}{\scriptsize{best pair}}
      \psfrag{combo002_componentsjet0ak07chargedcount_radialmomentak041Pt4vy}{\scriptsize{best pair}}
      \psfrag{components_jet0_ak05_charged_count_T_radial_moment_ak05_1_Pt_4v_y}{\scriptsize{charge * girth}}
      \psfrag{shape_optimal_pT_R_jet0_ak02_particles}{\scriptsize{optimal kernel}}
      \psfrag{subjet_1stPt_over_jetPt_jet0_ak05_sub_ak010}{\scriptsize{1st subjet R=0.5}}
      \psfrag{subjet_2ndPt_over_1stPt_jet0_ak05_sub_ak010}{\scriptsize{2nd subjet R=0.5}}
      \psfrag{subjet_3rdPt_over_jetPt_jet0_ak05_sub_ak010}{\scriptsize{3rd subjet R=0.5}}
      \psfrag{components_jet0_ak05_sub_ak010_avg_kT}{\scriptsize{avg $k_T$ of R$_\mathrm{sub}$\!=0.1}}
      \psfrag{decluster_kT_jet0_ak05_kt010}{\scriptsize{decluster\,$k_T$\,R$_\mathrm{sub}$\!=0.1}}
      \psfrag{shape_integ_pT_R_jet0_ak07_particles_bin_20_of_38_0100}{\scriptsize{jet shape $\Psi(0.1)$}}
      \psfrag{Signal eff}{\hspace{-1in}Quark Jet Acceptance}
      \psfrag{LHC 200 : Background Rejection}{Gluon Rejection}
      \psfrag{LHC 200 : Bkg Rejection}{Gluon Rejection}
      \includegraphics[width=0.51\textwidth]{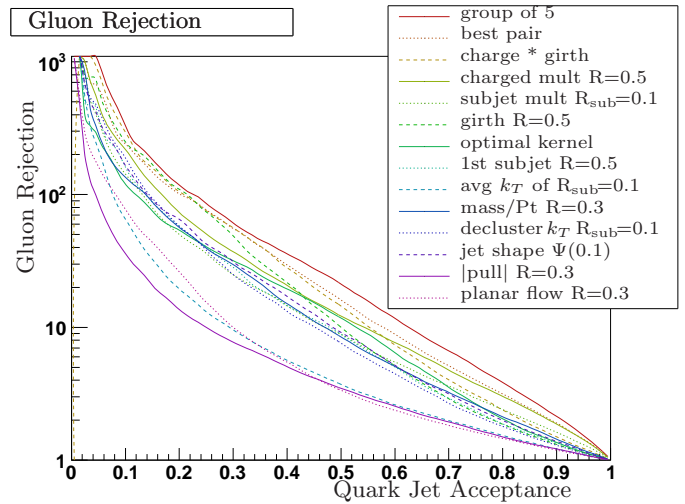}
\caption{
Gluon rejection curves for several observables as a function of
Quark Jet Acceptance.  The results for 200\,GeV Jets are shown, but other samples give similar results.
The best pair of observables is charged track multiplicity and linear radial moment (girth).
The best group of five also includes jet mass for the hardest subjet of size R=0.2,
the average $k_T$ of all R$_\mathrm{sub}$=0.1 subjets,
and the 3rd such small subjet's $p_T$ fraction.
} \label{fig:bkrej200}
\end{figure}

In summary, quite a number of single variables do comparably
well, while some (like pull or planar flow) do quite poorly at
gluon tagging. We examined many combinations of observables,
and found significant improvement by looking at pairs, but only
marginal gains beyond that. The results for the gluon rejection
as a function of quark efficiency are shown for a number of the
more interesting observables and combinations in
Figure~\ref{fig:bkrej200} for 200\,GeV jets. The relative
performance of variables changed little with $p_T$ even though
the optimal cuts do. Definitions and distributions of these
variables, and thousands of others, can be found on
\url{http://jets.physics.harvard.edu/qvg}. Good pairs of variables
included one from the discrete category described above, such
as particle count, and one more continuous shape variable, like
the linear radial moment (girth).

As an example using these curves to estimate the improvement in
a search's reach, consider $X \rightarrow WW \rightarrow q \bar
q q \bar q$ whose background is mostly 4-jets from QCD, each of
which is a gluon 80\% of the time \cite{Gallicchio:2011xc}. By
operating at 60\% quark efficiency, only 1/10$^\mathrm{th}$ of
gluons pass the tagger, which means $(20\%)^4$ of the total QCD
background passes. One measure of statistical significance in a
counting experiment is
$S/\sqrt{B}$, perhaps within a particular invariant mass
window. Any starting significance can be improved by a factor
of 3.2 using these cuts.  The 60\% operating point was chosen
to maximize this significance improvement for this particular
background composition, which highlights the need to
characterize background rejection for all signal efficiencies.

Measurements of these variables are underway, but it would be
very interesting to see distributions of and correlations
between as many of the variables in Figure~\ref{fig:bkrej200}
as possible. To this end, it has recently been observed that
99\% pure samples of quark jets can be obtained in
$\gamma$+2jet events, and 95\% pure samples of gluon jets can
be obtained in 3-jet events \cite{Gallicchio:2011xc}. These
samples could provide a direct evaluation of the tagging
technique at all jet $p_T$s, verify and help improve the Monte
Carlo generators, and provide a test of perturbative QCD.


The authors would like to thank Gavin Salam for early
consultation, the participants of the Boston Jet Physics
Workshop for useful feedback, the FAS Research Computing Group
at Harvard University and the DOE under Grant
DE-AC02-76CH03000, for support.

\end{document}